\documentclass[letter,traditabstract]{aa}

\usepackage{natbib}
\usepackage{amsmath,amssymb,verbatim}
\usepackage{graphicx}
\usepackage{hyperref}
\usepackage{txfonts}
\usepackage{color}

\newcommand{\rom}[1]{\MakeUppercase{\romannumeral #1}}

\begin{document}
	
	\title{
	Constraints on dark matter self-interaction \\ from the internal density profiles of X-COP galaxy clusters
	}
	
	\author{D. Eckert\inst{1} \and S. Ettori\inst{2,3} \and A. Robertson\inst{4} \and R. Massey\inst{4} \and E. Pointecouteau\inst{5} \and D. Harvey\inst{6} \and I. G. McCarthy\inst{7} }
	\institute{
		Department of Astronomy, University of Geneva, Ch. d'Ecogia 16, CH-1290 Versoix, Switzerland\\
		\email{Dominique.Eckert@unige.ch}
		\and
		INAF, Osservatorio di Astrofisica e Scienza dello Spazio, via Piero Gobetti 93/3, 40129 Bologna, Italy
		\and
		INFN, Sezione di Bologna, viale Berti Pichat 6/2, I-40127 Bologna, Italy
        \and
		Institute for Computational Cosmology, Department of Physics, Durham University, South Road, Durham DH1 3LE, UK
		\and
		IRAP, Université de Toulouse, CNRS, CNES, UPS, Toulouse, France
		\and
		Lorentz Institute, Leiden University, Niels Bohrweg 2, Leiden, NL-2333 CA, The Netherlands
		\and
		Astrophysics Research Institute, Liverpool John Moores University, 146 Brownlow Hill, Liverpool, L3 5RF, UK
	}
	
	\abstract{
	The fundamental properties of the postulated dark matter (DM) affect the internal structure of gravitationally-bound structures. In the cold dark matter paradigm, DM particles interact only via gravity. Their distribution is well represented by an Einasto profile with shape parameter $\alpha\approx0.18$, in the smallest dwarf galaxies or the most massive galaxy clusters alike. Conversely, if dark matter particles self-interact via additional forces, we expect the mass density profiles of DM halos to flatten in their central regions, thereby increasing the Einasto shape parameter. We measure the structural properties of the 12 massive X-COP galaxy clusters from observations of their hot gaseous atmosphere using the X-ray observatory \emph{XMM-Newton}, and of the Sunyaev-Zel’dovich effect using the \emph{Planck} all-sky survey. After removing morphologically disturbed systems, we measure Einasto shape parameters with mean $\langle\alpha\rangle = 0.19 \pm 0.03$ and intrinsic scatter $\sigma_{\alpha}=0.06$, in close agreement with the prediction of the cold dark matter paradigm. We use cosmological hydrodynamical simulations of cluster formation with self-interacting DM (\texttt{BAHAMAS-SIDM}) to determine how the Einasto shape parameter depends on the self-interaction cross section. We use the fitted relation to turn our measurements of $\alpha$ into constraints on the self-interaction cross section, which imply $\sigma/m < 0.19$ cm$^2$/g (95\% confidence level) at collision velocity $v_\mathrm{DM-DM}\sim1,000$ km/s. This is lower than the interaction cross-section required for dark matter self-interactions to solve the core-cusp problem in dwarf spheroidal galaxies, unless the cross section is a strong function of velocity.

	}
	
	\keywords{Cosmology: dark matter - X-rays: galaxies: clusters - Galaxies: clusters: general - Galaxies: clusters: intracluster medium}
	\maketitle

\section{Introduction}

In the currently favoured $\Lambda$ Cold Dark Matter ($\Lambda$CDM) cosmological paradigm, gravitationally-bound structures form hierarchically through the merging of smaller structures and accretion of material from the large-scale structure. The universality of the structure formation process should leave its imprint in the internal structure of DM halos, such that the density profiles of all halos from dwarf galaxies to galaxy clusters should follow a universal Navarro-Frenk-White shape \citep[NFW,][]{nfw96,nfw97} in which dark matter forms a central cusp with $\rho(r)\propto r^{-1}$ in the core, gradually steepening towards the outskirts asympotically as $\rho(r)\propto r^{-3}$. Recent N-body simulations have shown that the \citet{einasto65} functional form, in which the density profile is represented as a rolling power-law index $d\ln\rho/d\ln r \propto r^{-\alpha}$, provides a better representation of the shape of simulated halos \citep{navarro04,klypin16,ludlow16,brown20,ragagnin20}. The Einasto index $\alpha$ regulates the curvature of the profile, with the model profiles getting progressively more curved with increasing $\alpha$. In the CDM framework, the value of $\alpha$ is set primarily by the slope $n_s$ of the primordial power spectrum \citep{ludlow17,brown20}, corresponding to $\alpha\sim0.18$ for $n_s=0.96$ \citep{planck15_13}. The value of $\alpha$ should be close to universal in all halos, with the exception of a dependence on the ``peak height'', i.e. the amplitude of a local density fluctuation relative to its surroundings \citep{ludlow17}. 

Given the universality of the structural properties of DM halos in $\Lambda$CDM, deviations from the expected universal shape can be used to set constraints on the nature of DM. In particular, a non-vanishing DM self-interaction probability could impact the internal structure of halos, as collisional processes would transport heat across high-density regions, thereby homogenising the mass distribution and flattening the observed profiles in the cores of halos \citep{yoshida00,rocha13,robertson19,robertson20}. Self-interacting dark matter (SIDM) was proposed as a possible to the core-cusp problem in dwarf spheroidal galaxies \citep[see][for a review]{tulin18}. SIDM cross sections in the range $\sigma/m\sim1-5$ cm$^2$/g would produce central densities in broad agreement with the values observed in dwarf galaxies \citep[e.g.][]{spergel00,dave01,2018NatAs...2..907V,2019PhRvX...9c1020R}. 

On galaxy cluster scales, constraints on the SIDM cross section were obtained using dissociative cluster mergers \citep{randall08,bradac08,kahlhoefer14,gastaldello14,harvey15,robertson17}, strong gravitational lensing \citep{meneghetti01,sagunski21,andrade22}, and the wobbling of brightest central galaxies \citep{harvey19}. All the aforementioned studies concluded that the behaviour of DM in galaxy clusters is consistent with the collisionless scenario, with typical upper limits on the SIDM cross section at the level of $\sigma/m\lesssim 0.3-1$ cm$^2$/g. 

In this Letter, we provide high-precision estimates of the Einasto shape parameter $\alpha$ using the data of the \emph{XMM-Newton} Cluster Outskirts Project \citep[X-COP,][]{xcop}, a very large programme on \emph{XMM-Newton} providing a deep X-ray mapping of a sample of 12 massive galaxy clusters selected from the \emph{Planck} Sunyaev-Zel'dovich (SZ) survey. In a companion paper (Eckert et al. 2022, hereafter Paper \rom{1}), we presented a general framework to recover the mass profiles of galaxy clusters from X-ray and SZ data under the assumption that the gas is in hydrostatic equilibrium within the DM potential well. Here we apply our framework to recover the parameters of the Einasto functional form by fitting the X-COP data over a wide range in radius, which we then use to search for the effect of DM self-interaction. Throughout the paper, we assume a \emph{Planck} 2015 $\Lambda$CDM cosmology with $\Omega_m=0.308$, $\Omega_\Lambda=0.692$ and $H_0=67.8$ km/s/Mpc \citep{planck15_13}.

\section{Data analysis}

\subsection{The sample}

The X-COP sample \citep{xcop} consists of a set of 12 massive galaxy clusters in the redshift range $0.04<z<0.1$. The sample contains the most significant SZ detections from the \emph{Planck} PSZ1 catalog \citep{psz1} with the exception of three highly perturbed systems for which the validity of a radially-averaged analysis cannot be guaranteed. For each cluster, a minimum of 5 \emph{XMM-Newton} pointings are available to cover uniformly the gas distribution at least out to $2\times R_{500}$. 

The \emph{XMM-Newton} data were reduced in a homogeneous way \citep{ghirardini19,eckert19,ettori19} and the reduced data and thermodynamic profiles (gas density and spectroscopic temperature) are publicly available\footnote{\href{https://dominiqueeckert.wixsite.com/xcop}{https://dominiqueeckert.wixsite.com/xcop}}. In addition, \emph{Planck} SZ pressure profiles were derived for all systems following \citet{planck5} from the SZ signal reconstructed using the MILCA algorithm \citep{hurier13}. For more details on the X-COP program and on the \emph{XMM-Newton} and \emph{Planck} data reduction, we refer the reader to \citet{ghirardini19}.

\subsection{Hydrostatic mass reconstruction}
\label{sec:hydro}

To determine the total mass profile, we assume that the gas is fully thermalised within the gravitational well set by the DM, such that the pressure gradient locally balances the gravitational force,

\begin{equation}
	\frac{dP_{\rm gas}}{dr} = - \rho_{\rm gas}\frac{GM(<r)}{r^2}\label{eq:hse}.
\end{equation}

The validity of the assumption of hydrostatic equilibrium (HSE) in the case of the X-COP clusters was discussed in detail in \citet{ettori19,eckert19} and in Paper \rom{1}. With the exception of one peculiar system for which there is clear evidence of a breakdown of hydrostatic equilibrium \citep[A2319,][]{ghirardini18}, the comparison of HSE masses with other techniques \citep{ettori19} and the fitted gas and baryon fractions \citep{eckert19} indicate that the HSE assumption is valid at the $\lesssim20\%$ level out to $R_{200}$.

The methodology adopted to recover the mass profiles from X-ray surface brightness and spectroscopic temperature profiles and SZ pressure profiles is presented in detail in Paper \rom{1}. The code used to reconstruct the mass profiles from the X-ray and SZ data includes a wide range of effects (non-parametric gas density deprojection, PSF convolution, spectroscopic temperature weights, and profile covariance matrices) into an efficient Bayesian optimisation framework. The code is publicly available in the form of the Python package \texttt{hydromass}\footnote{\href{https://github.com/domeckert/hydromass}{https://github.com/domeckert/hydromass}}. The accuracy of the reconstruction technique was tested on mock X-ray observations of a synthetic NFW cluster in hydrostatic equilibrium, and the code was found to reproduce the input mass profile with an accuracy better than 3\%. 

In this paper, we focus on the reconstruction of Einasto mass model parameters. Namely, we describe the mass density profile using the \citet{einasto65} parametric form,

\begin{equation}
\rho_{\rm Einasto}(r) = \rho_s \exp\left[-\frac{2}{\alpha} \left( \left( \frac{r}{ r_s}\right)^\alpha - 1\right) \right].
\label{eq:einasto}
\end{equation}

The density profile is integrated numerically over the volume to determine the cumulative model mass as a function of radius, $M_{\rm mod}(<r)$. The gas density profile is described as a linear combination of King functions \citep{eckert20} and fitted jointly to the X-ray surface brightness profile. The hydrostatic equilibrium equation Eq. \ref{eq:hse} is then integrated to predict the 3D pressure profile,

\begin{equation}
P_{3D}(r) = P_0 + \int_{r}^{r_0} \rho_{\rm gas} \frac{G M_{\rm mod}(<r^\prime)}{r^{\prime 2}}\,dr^\prime\label{eq:integral}
\end{equation}

\begin{figure*}
	\resizebox{\hsize}{!}{\includegraphics[width=0.55\textwidth]{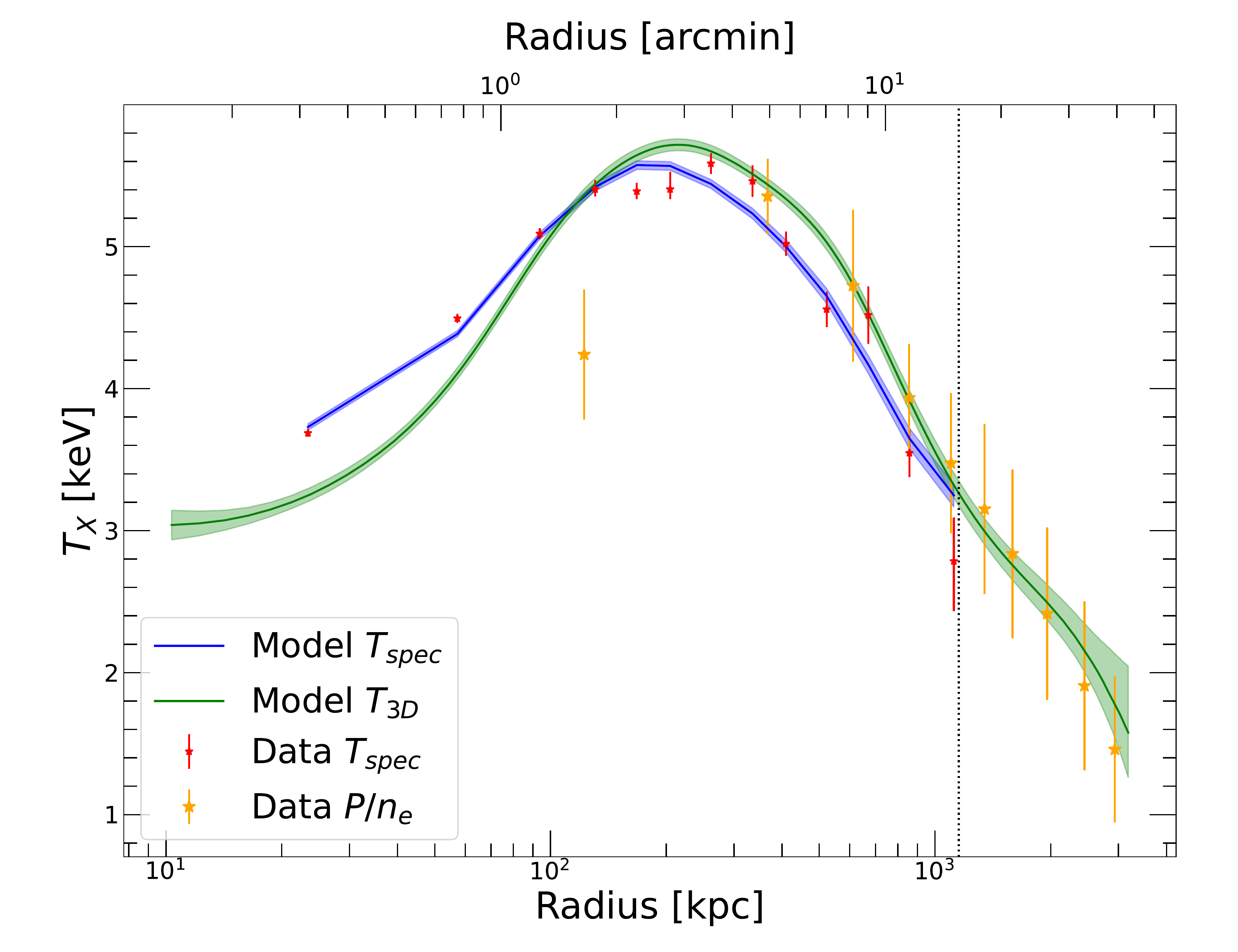}\includegraphics[width=0.45\textwidth]{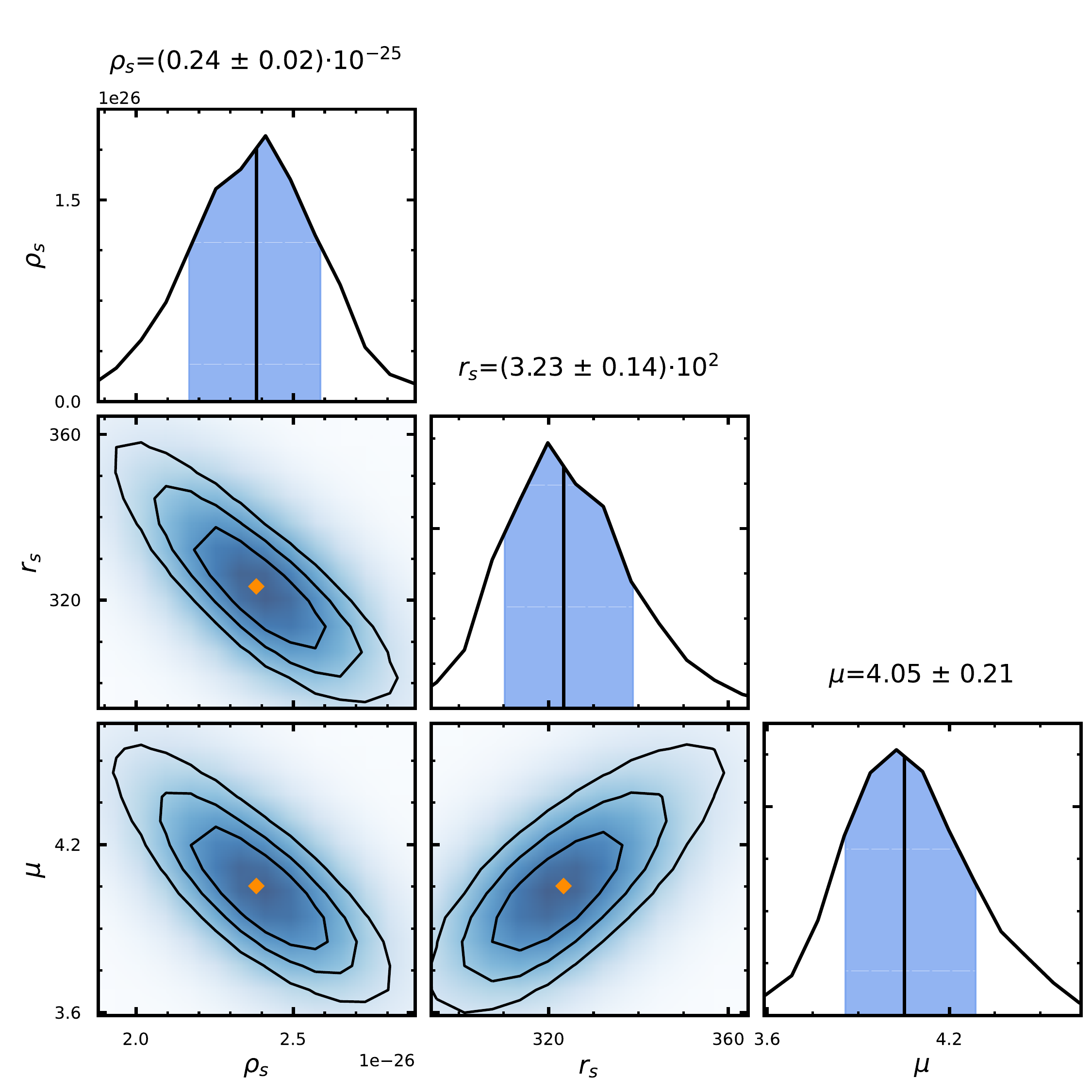}}
	\caption{\label{fig:a1795}Example Einasto model fit for A1795. The left-hand panel shows the spectroscopic X-ray temperatures (red points) and X/SZ temperatures obtained by dividing the SZ pressure by the gas density (orange stars). The projected spectroscopic-like model fitted to the X-ray temperatures is shown by the blue curve, whereas the 3D model temperature profile is shown as the green curve. The dotted vertical line indicates the fitted location of $R_{500}$. In the right-hand panel we show the posterior distributions of the fitted Einasto model parameters and the covariance between them.}
\end{figure*}

\noindent with $r_0$ the outermost radius of the SZ pressure profile and $P_0\equiv P(r_0)$ the integration constant corresponding to the pressure at the outer boundary. Finally, the model pressure is projected along the line of sight and convolved with the instrumental PSF to predict the expected spectroscopic temperature profile. For more details on the mass reconstruction technique and a thorough discussion of the associated systematic uncertainties, we refer the reader to Paper \rom{1}.

Following \citet{ettori19}, rather than fitting for the characteristic density $\rho_s$ we optimise for the unitless normalisation $c$, which is related to $\rho_s$ as 

\begin{equation}
\rho_s =  \frac{\Delta}{3} \frac{c^3}{\log(1 + c) - c / (1 + c)}  \rho_c(z) 
\end{equation}

\noindent with $\Delta=200$ the chosen fit overdensity and $\rho_c(z) = 3H(z)^2 / 8\pi G$ the critical density of the Universe. In addition, to enhance the stability of the procedure we optimise for $\mu\equiv1/\alpha$ \citep{mamon05}. All together, our model includes four free parameters: the scale radius $r_{s}$, the normalisation $c$, the inverse $\mu$ of the Einasto shape parameter $\alpha$, and the integration constant $P_0$. Weak Gaussian priors are set on the Einasto model parameters (see Table \ref{tab:priors}), whereas for $P_0$ we set a Gaussian prior with mean and standard deviation equal to the outermost SZ pressure value and its uncertainty. The adopted priors are described in detail in Table \ref{tab:priors}. In every case, the posterior distributions are much narrower than the prior and clearly pull away from it, therefore the results of this paper weakly depend on the choice of the prior. Using uniform priors over the range given in Table \ref{tab:priors} also returns similar results and uncertainties for all parameters.

\begin{table}
\caption{\label{tab:priors}Normal priors on the Einasto fit parameters. Here $P_{m}$ and $dP_{m}$ denote the outermost SZ pressure value and its error.}
\begin{tabular}{ccccc}
\hline
Parameter & Mean & $\sigma$ & Min & Max \\
\hline
\hline
$r_{s}$ [kpc] & 700 & 300 & 100 & 3000 \\
$c$ & 1.8 & 1.5 & 0 & 10\\
$\mu$ & 5 & 3 & 0.2 & 20\\
$P_0$ & $P_{m}$ & $dP_{m}$ & $P_{m}-2dP_{m}$ & $P_{m}+2dP_{m}$\\
\hline
\end{tabular}

\end{table}

The optimisation is performed using the No U-Turn Sampler (NUTS) as implemented in \texttt{PyMC3} \citep{pymc3}. Four independent chains are run in parallel, each with 1,000 tuning steps and 1,000 output samples. In Paper \rom{1} we showed that the Einasto model provides a good representation of the data at hand and closely follows the results obtained with a non-parametric technique that makes no assumption on the shape of the mass profile. As an example, in Fig. \ref{fig:a1795} we show the result of the fitting procedure for A1795. The left-hand panel shows the X-ray spectroscopic temperature profile and the X/SZ temperature profile obtained by dividing the SZ pressure by the X-ray gas density. The best-fit 3D and projected model and the corresponding $1\sigma$ error are displayed as well. The data are very well represented by the model throughout the entire radial range. We note that constraints over a wide radial range are required to break the degeneracy between the model parameters and determine the Einasto shape parameter with good accuracy. In Appendix \ref{sec:hse_alpha} we quantify the impact of hydrostatic bias on the recovered Einasto indices and show that the resulting values of $\alpha$ are mildly affected by deviations from HSE. Moreover, we note that while most galaxy clusters should exhibit an elliptical shape, our results are mildly affected by the assumption of spherical symmetry \citep{buote12}. 

\section{Results}

\subsection{Einasto shape parameter}

\begin{figure}
	\resizebox{\hsize}{!}{\includegraphics{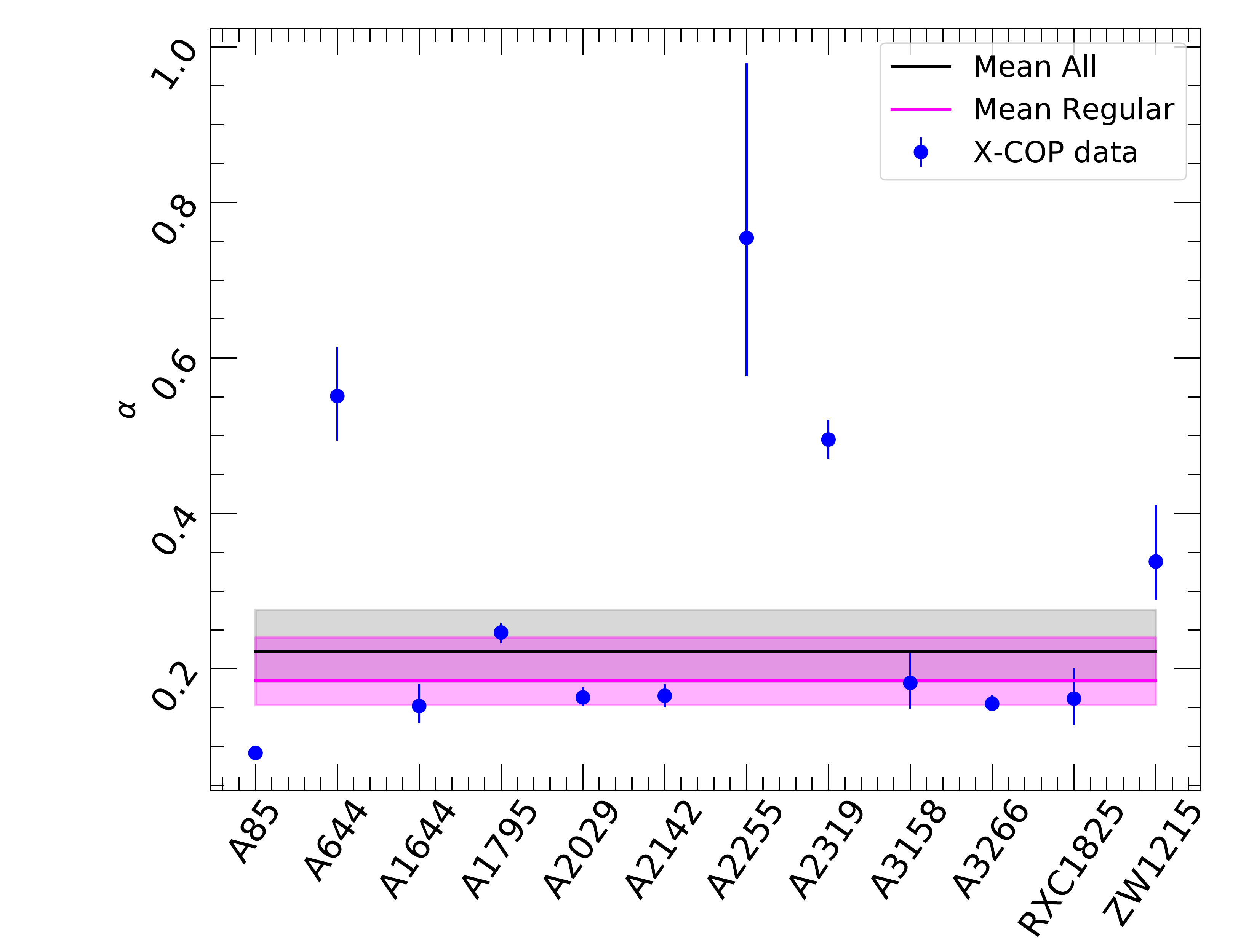}}
	\caption{\label{fig:alpha}Measurements of the Einasto shape parameter $\alpha$ for the 12 X-COP clusters. The black line and grey shaded area show the fitted mean value and its error, whereas the magenta curve shows the fit to the ``unperturbed'' subsample with $w<0.0015$.}
\end{figure}

We applied our mass profile reconstruction technique to all 12 X-COP systems. The resulting parameters are provided in Table \ref{tab:params}. As can be seen in Table \ref{tab:params}, in spite of the substantial covariance between the parameters the data at hand are of sufficient quality to determine the Einasto shape parameter $\alpha$ (or rather its inverse $\mu$) with good precision. For each system, we used the posterior distribution of $\mu$ to determine the corresponding value of $\alpha$ and its uncertainty. In Fig. \ref{fig:alpha} we show the estimated values of $\alpha$ for all the systems. In the majority of cases, the measured values of $\alpha$ are in the range 0.15-0.3, with three systems (A2319, A644, and A2255) exhibiting larger values of $\alpha$ ($\gtrsim 0.5$). 

\begin{table*}
	\begin{center}
	\caption{\label{tab:params}Result of Einasto fitting procedure for the 12 X-COP clusters }
	\begin{tabular}{lccccccc}
\hline
Cluster & $M_{500,Einasto}$ & $r_{s}$ & $\rho_{s}$ & $\mu$ & $w$ & $R_{\rm min}$ & $R_{\rm max}$\\ 
 & $10^{14} M_\odot$ & kpc & $10^{-27}$ g/cm$^3$ &  & $10^{-3}$ & $R_{200}$ & $R_{200}$ \\ 
\hline
\hline
A85 & $6.35_{-0.08}^{+0.08}$ & $1688_{-153}^{+160}$ & $1.08_{-0.16}^{+0.21}$ &  $10.88_{-0.41}^{+0.41}$ & $3.85\pm0.02$ & 0.009 & 1.03 \\ 
A644 & $6.06_{-0.30}^{+0.30}$ & $334_{-12}^{+13}$ & $35.86_{-3.15}^{+3.17}$ &  $1.82_{-0.19}^{+0.21}$ & $20.97\pm0.15$ & 0.013 & 0.87 \\ 
A1644 & $2.97_{-0.09}^{+0.09}$ & $982_{-174}^{+238}$ & $1.94_{-0.65}^{+0.87}$ &  $6.57_{-1.03}^{+1.12}$ & $13.76\pm0.08$ & 0.010 & 1.30 \\ 
A1795 & $4.36_{-0.10}^{+0.12}$ & $323_{-13}^{+16}$ & $23.82_{-2.16}^{+2.06}$ &  $4.05_{-0.20}^{+0.24}$ & $2.36\pm0.02$ & 0.012 & 1.29 \\ 
A2029 & $7.76_{-0.22}^{+0.21}$ & $488_{-39}^{+36}$ & $14.18_{-1.84}^{+2.52}$ &  $6.12_{-0.44}^{+0.43}$ & $0.85\pm0.01$ & 0.012 & 1.13 \\ 
A2142 & $9.41_{-0.19}^{+0.20}$ & $1003_{-103}^{+147}$ & $3.89_{-0.88}^{+0.89}$ &  $6.04_{-0.49}^{+0.58}$ & $4.51\pm0.03$ & 0.007 & 1.07 \\ 
A2255 & $5.58_{-0.27}^{+0.29}$ & $699_{-39}^{+49}$ & $6.62_{-1.02}^{+1.07}$ &  $1.33_{-0.30}^{+0.41}$ & $31.40\pm0.26$ & 0.028 & 1.24 \\ 
A2319 & $7.79_{-0.10}^{+0.09}$ & $437_{-6}^{+7}$ & $21.77_{-0.82}^{+0.78}$ &  $2.02_{-0.10}^{+0.11}$ & $33.08\pm0.17$ & 0.020 & 1.12 \\ 
A3158 & $4.32_{-0.15}^{+0.14}$ & $730_{-129}^{+176}$ & $4.34_{-1.47}^{+1.99}$ &  $5.49_{-0.99}^{+1.23}$ & $5.87\pm0.04$ & 0.011 & 1.06 \\ 
A3266 & $7.44_{-0.13}^{+0.14}$ & $1528_{-132}^{+81}$ & $1.53_{-0.14}^{+0.28}$ &  $6.45_{-0.44}^{+0.40}$ & $30.84\pm0.16$ & 0.009 & 0.92 \\ 
RXC1825 & $3.93_{-0.09}^{+0.09}$ & $579_{-74}^{+135}$ & $6.34_{-2.18}^{+2.08}$ &  $6.18_{-1.22}^{+1.68}$ & $8.03\pm0.06$ & 0.013 & 1.19 \\ 
ZW1215 & $6.65_{-0.32}^{+0.30}$ & $689_{-81}^{+85}$ & $6.86_{-1.39}^{+1.97}$ &  $2.96_{-0.52}^{+0.50}$ & $3.73\pm0.03$ & 0.013 & 0.94 \\ 
\hline
\end{tabular}

	\end{center}
\textbf{Column description}: Cluster name; Value of $M_{500}$ and uncertainty from the best-fit Einasto model; Parameter values and uncertainties for the Einasto parameters $r_s$, $\rho_s$, and $\mu=1/\alpha$ (see Eq. 
\ref{eq:einasto}); Centroid shift $w$ (see Eq. \ref{eq:w}); Lower and upper bounds of the fitting range as a fraction of $R_{200}$.
\end{table*}

To determine the average value of $\alpha$ and its intrinsic scatter across the population, we describe the population with a constant mean value and a normal intrinsic scatter. The total observed scatter is assumed to be the quadratic sum of the intrinsic and the statistical scatter. We then fit the data in \texttt{PyMC3} using this simple model and recover output chains for the mean value of $\alpha$ and the intrinsic scatter. Through this procedure, we found a mean value $\langle\alpha\rangle=0.22\pm0.04$, with a substantial intrinsic scatter $\sigma_{\alpha}=0.14_{-0.05}^{+0.10}$. The majority of the scatter can be attributed to the three outliers quoted above (A2319, A644, A2255). Incidentally, in \citet{ettori19} the mass profiles of these same systems were found to be better represented by models exhibiting a central core (Burkert, non-singular isothermal sphere). A closer look at these systems shows that they are all unrelaxed systems with large deviations from spherical symmetry. To investigate this point, we used the [0.7-1.2] keV mosaic images of our systems to measure their centroid shift $w$, which quantifies the change in the position of the X-ray centroid as a function of the used aperture. Namely, we define $N$ apertures of decreasing radius $R_i$, with $R_1=R_{500}$, and determine the X-ray centroid $\Delta_i$ within each of them. The centroid shift is then defined as

\begin{equation}
	w = \frac{1}{R_{500}} \sqrt{\frac{1}{N-1}\sum_{i=1}^N (\Delta_i - \Delta_{500})^2}
	\label{eq:w}
\end{equation}

\noindent with $\Delta_i-\Delta_{500}$ the difference in the centroid position between aperture $R_i$ and $R_{500}$. The centroid shift is known to be an accurate indicator of a cluster's dynamical state \citep{rasia13}, i.e. the more relaxed a cluster is, the smaller the value of $w$. The values of $w$ for all X-COP clusters are given in Table \ref{tab:params}. 

Most studies \citep[e.g.][]{ohara06,cassano,weissmann13} place the boundary between relaxed and unrelaxed systems in the range $w=0.01-0.02$. We can see that all the quoted outliers exhibit values of $w$ greater than $0.02$, which shows that their X-ray morphology is unrelaxed. A number of possible issues may render our mass reconstruction unreliable in unrelaxed systems. First, the validity of the hydrostatic equilibrium assumption in unrelaxed systems is unclear, as these systems are likely undergoing merging events at the present time. Second, the position of the X-ray centroid may not coincide with the bottom of the potential well, as the gas experiences hydrodynamical effects such as ram pressure that may displace it from the underlying DM distribution. In case our profiles are miscentered with respect to the bottom of the potential well, we expect the reconstructed mass profiles to be substantially flatter than the true profiles, which would bias the reconstructed values of $\alpha$ towards high values. 

To avoid the potential impact of miscentering and hydrostatic bias, we selected a subsample of X-ray regular systems with $w<0.015$. This subsample excludes four objects with high centroid shifts, for which the offset between the X-ray peak and the bottom of the gravitational potential may be large and thus the value of $\alpha$ may be unreliable. We fitted again the mean value of $\alpha$ and its intrinsic scatter in the regular subsample, and found that the mean value slightly decreases, $\langle\alpha_{\rm rel}\rangle=0.19\pm0.03$. The scatter in the population decreases substantially compared to the full population to $\sigma_{\alpha, rel}=0.06_{-0.03}^{+0.04}$, showing that the systems for which reliable measurements of $\alpha$ can be made exhibit very similar values of the Einasto shape parameter. 

The average value of $\alpha$ in the regular subsample is significantly lower than the value of $0.29\pm0.04$ reported by \citet{mantz16} using \emph{Chandra} data in a sample of 40 relaxed clusters. The difference may arise from the radial range probed in the two studies, as in most cases the data used by \citet{mantz16} do not extend beyond $\sim R_{2500}$. Alternatively, the systems used by \citet{mantz16} may have larger peak heights than the X-COP clusters, as the Einasto parameter should be a strong function of the peak height \citep{klypin16}. Conversely, our results agree with the value of $0.19\pm0.07$ measured by \citet{umetsu14} on the stacked weak lensing shear profile of the CLASH sample, although our constraints are substantially tighter. 

\subsection{Dependence of $\alpha_{Einasto}$ on SIDM cross section}

\begin{figure*}
    \resizebox{\hsize}{!}{
    \includegraphics{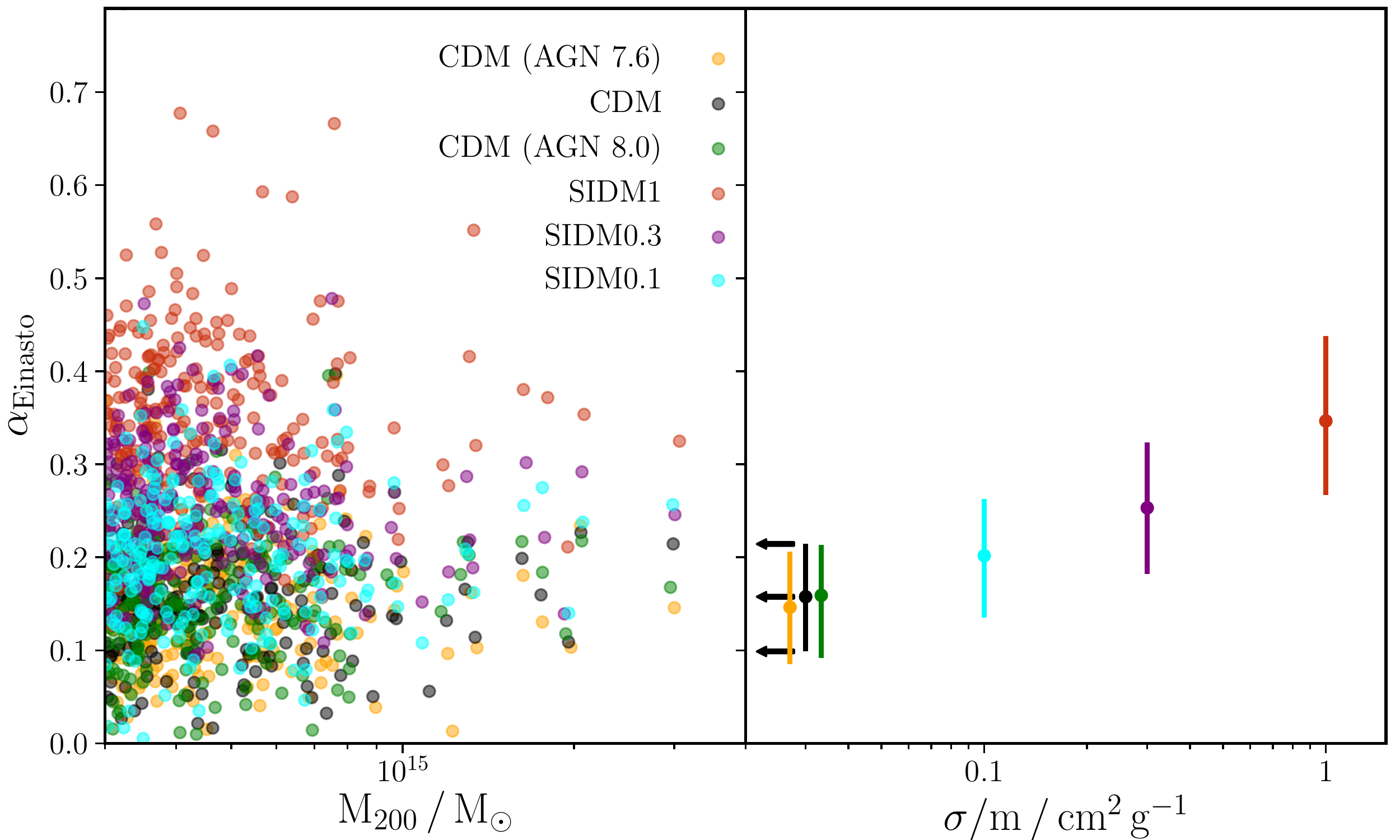}}
    \caption{Simulations of the structural properties of massive DM halos in the CDM and SIDM frameworks. The left-hand panel shows the Einasto shape parameter $\alpha_{\rm Einasto}$ for individual simulated halos with varying SIDM cross section (0.1 cm$^2$/g, cyan; 0.3, purple; and 1.0, red; \citealt{robertson20}). For comparison, we also show the effect of varying baryonic physics (reference model, black; models with lower (yellow) and higher (green) feedback efficiency; \citealt{mccarthy17}). The right-hand panel shows the median and $1\sigma$ percentiles of $\alpha_{\rm Einasto}$ values for the various simulations runs.} 
    \label{fig:alpha_sims}
\end{figure*}

To find how $\alpha_{\rm Einasto}$ is expected to depend on SIDM cross-section, we use the \texttt{BAHAMAS-SIDM} cosmological simulations \citep{mccarthy17,robertson20}. These simulate the growth of structure in a large volume of the universe ($400\,h^{-1}$\,Mpc on a side), which includes approximately 230 clusters with mass $M_{200}>3\times10^{14}\,M_\odot$ by redshift $z=0$. 
Four separate simulations start from identical initial conditions, then model SIDM with velocity-independent cross section $\sigma/m=0$\,cm$^2$/g (CDM), $0.1$\,cm$^2$/g (SIDM0.1), $0.3$\,cm$^2$/g (SIDM0.3), and $1.0$\,cm$^2$/g (SIDM1). During each time-step, SIDM particles have a small chance of elastically scattering off nearby neighbours \citep{robertson19}. 
The evolution of baryonic particles uses the \texttt{BAHAMAS} model \citep{mccarthy17}, which includes a wide range of physical processes for galaxy evolution (cooling, star formation, stellar and AGN feedback) and reproduces both the galaxy stellar mass function and X-ray scaling relations. 

We fitted an Einasto profile to the distribution of total mass (DM + baryons) for all haloes with $M_{200} > 3\times10^{14}\,M_\odot$ in each simulation's $z=0$ snapshot. The spherically-averaged density profile of each simulated halo was calculated about the most-bound particle, and the fit was done by minimising the sum of squared residuals in $\log\rho$ over 40 radial bins, logarithmically spaced between $0.01 \times R_{200}$ and $R_{200}$. The adopted radial range matches well with the range over which the observational constraints are obtained (see Table \ref{tab:params}). 
The Einasto model was found to provide a good representation of the mass profiles, although the isothermal Jeans model proposed by \citet{2014PhRvL.113b1302K} and tested on simulated halos by \citet{robertson20} describes the simulated halos more accurately. Figure~\ref{fig:alpha_sims} shows the best-fit value of $\alpha_{\rm Einasto}$ for each simulated cluster, as well as the median and 16th to 84th percentiles for different SIDM cross-sections. The median value of $\alpha_{\rm Einasto}$ increases from  0.18 for CDM \citep[as previously known;][]{navarro04,ludlow16} to 0.35 for SIDM1, demonstrating the shallower central slope of SIDM clusters. Increasing the mass threshold to $M_{200}>5\times 10^{14} M_\odot$ does not significantly change the average value of $\alpha$.

\begin{figure*}
	\resizebox{\hsize}{!}{\includegraphics[width=0.5\textwidth]{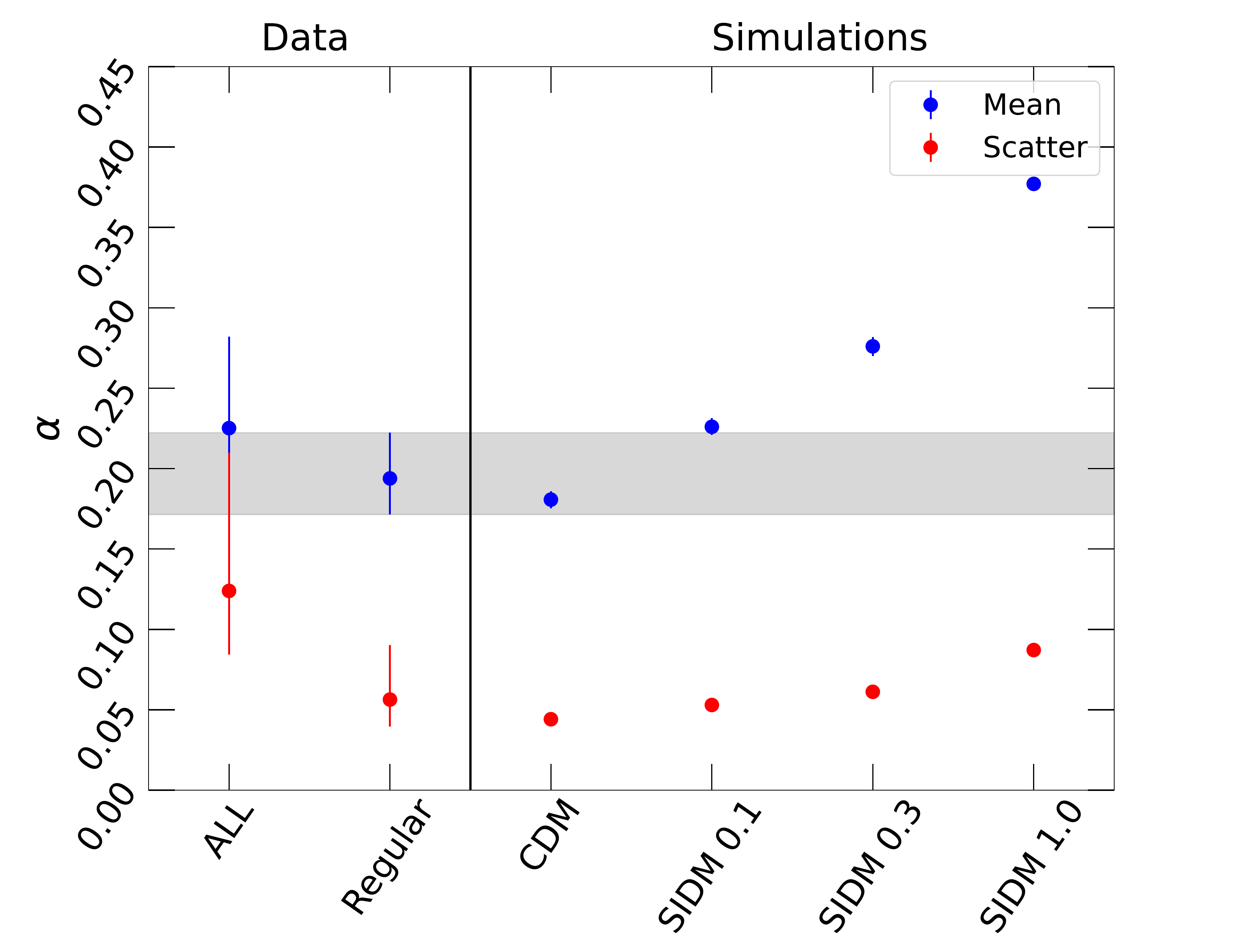}\includegraphics[width=0.5\textwidth]{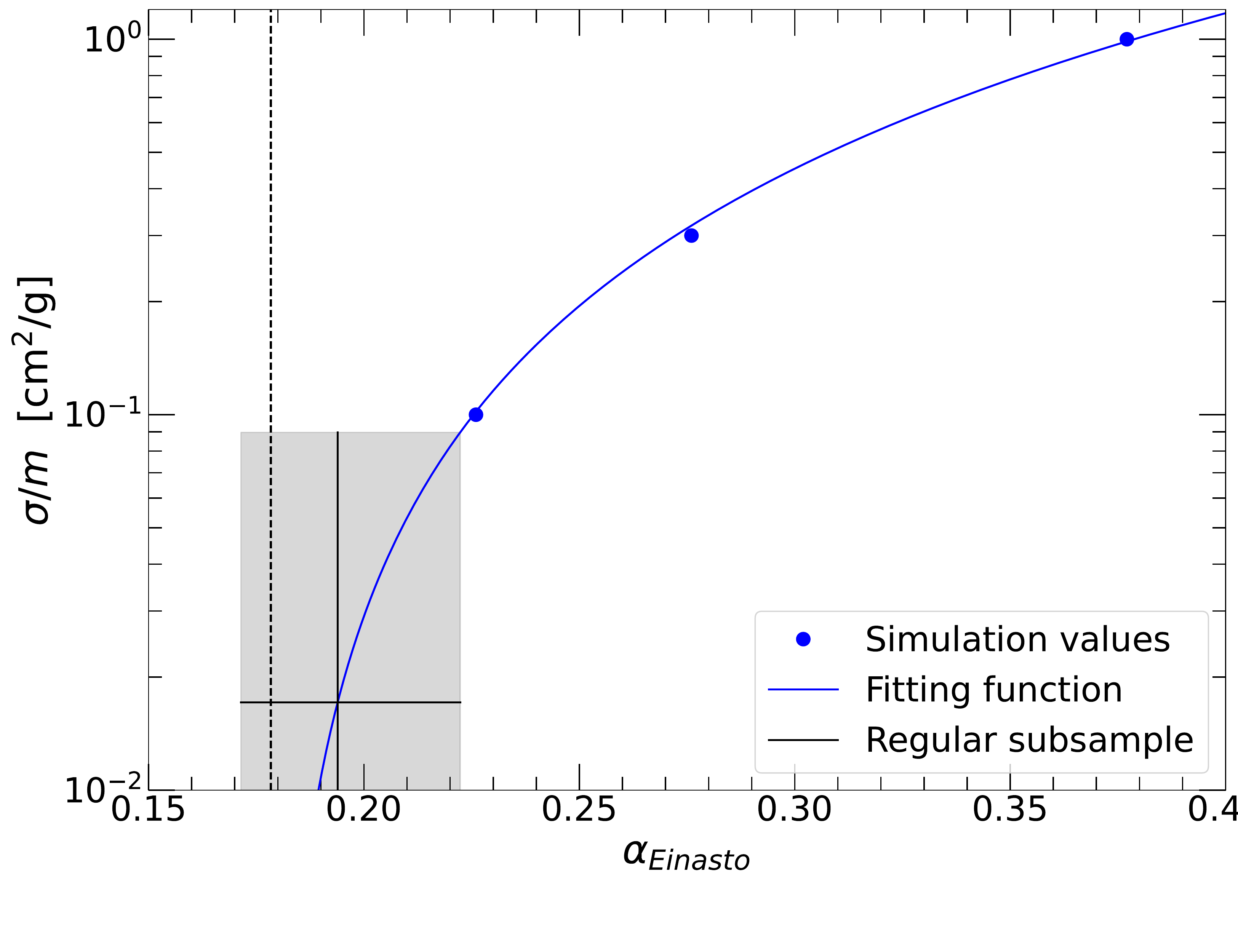}}
	\caption{Relation between the Einasto shape parameter $\alpha_{\rm Einasto}$ and the dark matter self-interaction cross section $\sigma/m$ for several sets of numerical simulations with varying $\sigma/m$ \citep{robertson20}. The left-hand panel shows the fitted mean values of $\alpha$ in the relaxed subsample for the CDM and SIDM cases, compared to the average value in the full X-COP sample (ALL) and in the regular X-ray subsample. In the right-hand panel we show the dependence of $\alpha$ on $\sigma/m$. The solid blue line shows a fit to the relation with a power law, whereas the dashed vertical line indicates the average value in CDM. The black line and grey shaded area show the mean value of $\alpha_{Einasto}$ and the $1\sigma$ uncertainty in the regular X-COP subsample, with the corresponding constraints on $\sigma/m$.}
	\label{fig:alpha_vs_sigma}
\end{figure*}

Since we are fitting the total density profile (including baryons), it is important to check whether the dependence of $\alpha$ on the SIDM cross section could be degenerate with uncertainties associated with the evolution of baryons. The main source of uncertainty in the baryonic model is the implementation of AGN feedback, which affects the star formation efficiency and the gas density profile. \texttt{BAHAMAS} implements the \citet{booth09} model, in which the central black holes store accreted energy until it exceeds a given threshold. The stored energy is then released thermally to reheat the surrounding medium. A low energy threshold provides gentle, continuous energy injection, whereas a high value of the energy threshold leads to burstier AGN feedback. The default \texttt{BAHAMAS} model implements the AGN 7.8 model, which was found to reproduce accurately the X-ray scaling relations and gas fractions in galaxy clusters and groups \citep{lebrun14,mccarthy17}. For comparison, Fig.~\ref{fig:alpha_sims} also shows the median value and scatter of $\alpha_{\rm Einasto}$ in CDM simulations with a decreased (AGN 7.6) or increased (AGN 8.0) energy threshold \citep{mccarthy18}, which bracket the range of parameters reproducing the observed scaling relations and gas fractions. The insignificant dependence on energy threshold demonstrates that $\alpha_{\rm Einasto}$ provides a clean test of DM self-interactions, regardless of the adopted prescription for baryonic physics.

\subsection{Constraints on SIDM cross section}

To test for deviations from the $\Lambda$CDM model, we compared our observed data against predictions of SIDM simulations. We used the values of $\alpha_{\rm Einasto}$ fitted to individual halos in each simulation set and applied the same procedure as for the observed halos, i.e.\ we fitted the data set with a constant mean value and a normal intrinsic scatter. In addition, to allow for a meaningful comparison with the regular X-COP subsample, we selected a subsample of relaxed systems according to the DM offset parameter $x_{\rm off}$, which determines the offset between the most bound particle and the center of mass of the halo, relative to the system's virial radius \citep{maccio07}. $x_{\rm off}$ can be viewed as a DM analog of the centroid shift, such that a selection of halos with low $x_{\rm off}$ should correspond to our selection of X-ray regular systems. We select a subsample of 114 relaxed systems with $x_{\rm off}<0.05$ to be compared with the regular X-COP subsample. We found that $\alpha_{\rm Einasto}$ slightly depends on $x_{\rm off}$, with the mean value of the CDM sample increasing from $0.16$ in the full population to $0.18$ in the low $x_{\rm off}$ subsample. We note that the effect goes in the opposite direction with respect to the observations, where restricting to the X-ray regular subsample slightly reduces the mean value of $\alpha_{\rm Einasto}$. We interpret this difference as resulting from miscentering in dynamically active systems, for which finding the bottom of the potential well is difficult, which flattens the mass profiles in the innermost regions and thus raises the value of $\alpha$. Conversely, numerical simulations are always able to pinpoint the most bound particle in 3D. To make a meaningful comparison on the full sample, applying the exact same analysis procedure to mock observations extracted from numerical simulations would be needed, which is beyond the scope of this Letter.

Given the observational uncertainty for the unrelaxed clusters, we focus here on the comparison of the X-ray regular subsample with the simulated low $x_{\rm off}$ sample. In the left-hand panel of Fig. \ref{fig:alpha_vs_sigma} we show the mean value of $\alpha$ and the intrinsic scatter in the low $x_{\rm off}$ subsample for the four simulations with varying $\sigma/m$. We compare the results with the mean values obtained for the full X-COP sample and the X-ray regular subsample. We can see that the fitted mean value for the regular subsample agrees very well with the CDM expectation, both in terms of the mean value and the intrinsic scatter. We can thus set an upper limit on $\sigma/m$ by comparing the mean value of $\alpha$ to the value expected in the various simulation sets. To interpolate between the discrete values of $\sigma/m$, we describe the relation between $\alpha$ and $\sigma/m$ as 

\begin{equation}
	\alpha_{\rm Einasto} = \alpha_0 + \alpha_1 \left( \frac{\sigma/m}{\mbox{1 cm$^2$/g}}\right)^\gamma\label{eq:alpha_vs_sigma}
\end{equation}

\noindent i.e.\ $\alpha_0$ represents the mean value of $\alpha$ in CDM and the dependence of $\alpha$ on $\sigma/m$ is described as a power law. For the low $x_{\rm off}$ subsample, we found $\alpha_0=0.178$, $\alpha_1=0.20$, and $\gamma=0.63$. This model accurately reproduces the data, as can be seen in the right-hand panel of Fig. \ref{fig:alpha_vs_sigma}, where the model curve is compared with the mean values of the simulation sets. 

To translate our constraints on $\alpha$ into upper limits on $\sigma/m$, we convert each value of $\langle\alpha_{\rm rel}\rangle$ in the output chain into a value of $\sigma/m$ through Eq. \ref{eq:alpha_vs_sigma} and construct a posterior distribution for $\sigma/m$. As expected, the posterior distribution is consistent with zero. We then compute a one-tailed 95\% upper limit as the 95th percentile of the posterior distribution, 
\begin{equation}
	\frac{\sigma}{m} < 0.19 \, {\rm cm}^2/{\rm g,~at~} v_{\mathrm{DM}}\! \approx\!1,000~\mbox{(95\% confidence level)}.
\end{equation}

Similar constraints on $\sigma/m$ are obtained if we use a different method to relate $\alpha$ to $\sigma/m$ (linear interpolation, cubic spline, or second-order polynomial), which shows that our result is robust to changes in the interpolation method. Moreover, the statistical uncertainties in the mean value of $\alpha$ within each simulation set are much smaller than the observational errors (see the left-hand panel of Fig. \ref{fig:alpha_vs_sigma}) and thus they can safely be ignored.

The upper limit quoted here is one of the most stringent to date. For comparison, offsets between DM and baryons in individual merging clusters imply $\sigma/m\lesssim 1$ cm$^2$/g \citep[e.g.][]{randall08} or $\sigma/m<0.47$ cm$^2$/g at 95\% confidence level from an ensemble of 72 merging systems \citep{harvey15}. The wobbling of brightest cluster galaxies around the centre of DM (which also probes the radial mass distribution) constrains $\sigma/m\lesssim0.39$ cm$^2$/g at 95\% confidence level \citep{harvey19}. Using a similar technique to that presented in this paper, but from the gravitational lensing profiles of 8 clusters, \citet{andrade22} found $\sigma/m\lesssim0.13$ cm$^2$/g at 95\% confidence level. Our results are of similar precision and, combined with the studies quoted above, imply that SIDM cross sections larger than $0.1$~cm$^2$/g would be in tension with existing data.

Removing the dark matter cusps in dwarf spheroidal galaxies, and solving the `core-cusp problem' \citep[see][for a review]{bullock17}, requires cross sections larger than 1~cm$^2$/g at $v_{\mathrm{DM}}\sim10$~km/s \citep{spergel00,dave01,lan20}. This appears impossible for SIDM models in which the cross-section is constant. However, many particle physics models predict a cross section that is a strong \emph{decreasing} function of velocity, $\propto v_{\mathrm{DM}}^{-4}$ above the (unknown) mass scale of the dark sector force mediator particle. In these models, SIDM may still be a viable solution to the core-cusp problem. 

Generally speaking, the technique developed here opens the possibility of probing SIDM at much higher precision using future surveys like \emph{eROSITA} \citep{erosita} and \emph{Euclid} \citep{euclid_clusters}, which will detect tens of thousands of clusters and will potentially improve the constraints on $\alpha_{\rm Einasto}$. The technique proposed here can also be applied to systems of lower mass, to directly test for variation in the SIDM cross section with velocity. 

\section{Conclusion}

We have measured 
the internal DM structure of galaxy clusters, and compared our observations with clusters in CDM and SIDM simulations. Our results can be summarised as follows:

\begin{itemize}
	\item The hydrostatic mass profiles of the 12 X-COP galaxy clusters are well represented by an Einasto model over more than two decades in radius. Data quality is sufficient to break the degeneracy between the various parameters of the Einasto model, and determine the Einasto index with high precision.
	
	\item Most X-COP clusters have Einasto index $0.15<\alpha<0.3$, with mean $\langle\alpha\rangle=0.22\pm0.04$. The substantial intrinsic scatter, $\sigma_{\alpha}=0.14_{-0.05}^{+0.10}$, is dominated by a few outliers with $\alpha\gtrsim 0.5$, all of which appear morphologically disturbed, as indicated by a large centroid shift. The high Einasto indices are therefore likely a systematic effect in the reconstruction of the mass profiles, caused by hydrostatic bias and/or miscentering. The subset of 8 X-ray regular systems with low centroid shift have mean $\langle\alpha\rangle=0.19\pm0.03$. Their intrinsic scatter,  $\sigma_\alpha=0.06_{-0.03}^{+0.05}$, is so small that $\alpha$ appears close to universal.
	
	\item In CDM and SIDM simulations, we find that cluster mass profiles become more curved as the SIDM cross section $\sigma/m$ increases, with $\alpha\approx 0.18 + 0.20 ( (\sigma/m)/(\mbox{1 cm$^2$/g)})^{0.63}$. This predicted trend is robust to changes in the adopted prescription for sub-grid baryonic processes.
	
	\item Comparing the relaxed subset of observed clusters with those in \texttt{BAHAMAS-SIDM} simulations allows us to set an upper limit $\sigma/m<0.19$ cm$^2$/g (95\% c.l.) on the cross-section for dark matter interactions at a collision velocity $v_{\mathrm{DM}}\approx1000$~km/s. This is one of the most stringent constraints to date, and substantially lower than the value required to explain the core-cusp problem in dwarf spheroidal galaxies, unless the cross section is a strong decreasing function of collision velocity. 
	
\end{itemize}
  
Future studies could apply our technique to large samples of galaxies, groups and clusters. It has potential to improve substantially the constraints on the fundamental self-interaction cross section of dark matter, or even to measure it as a function of velocity.

\bibliographystyle{aa}
\bibliography{einasto}

\appendix

\section{Impact of hydrostatic bias on the shape parameter}
\label{sec:hse_alpha}

\begin{figure*}
\resizebox{\hsize}{!}{\includegraphics[width=0.5\textwidth]{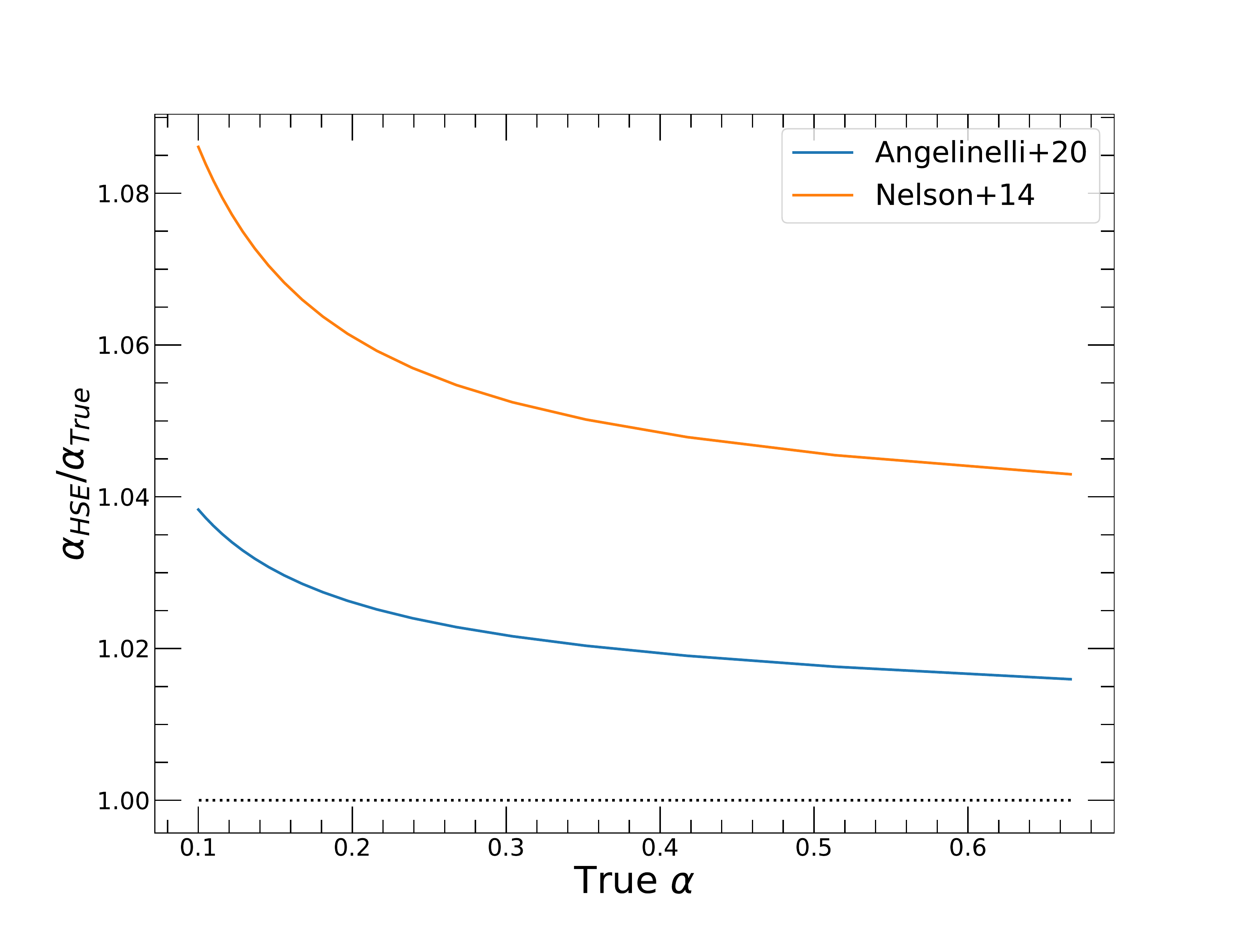} \includegraphics[width=0.5\textwidth]{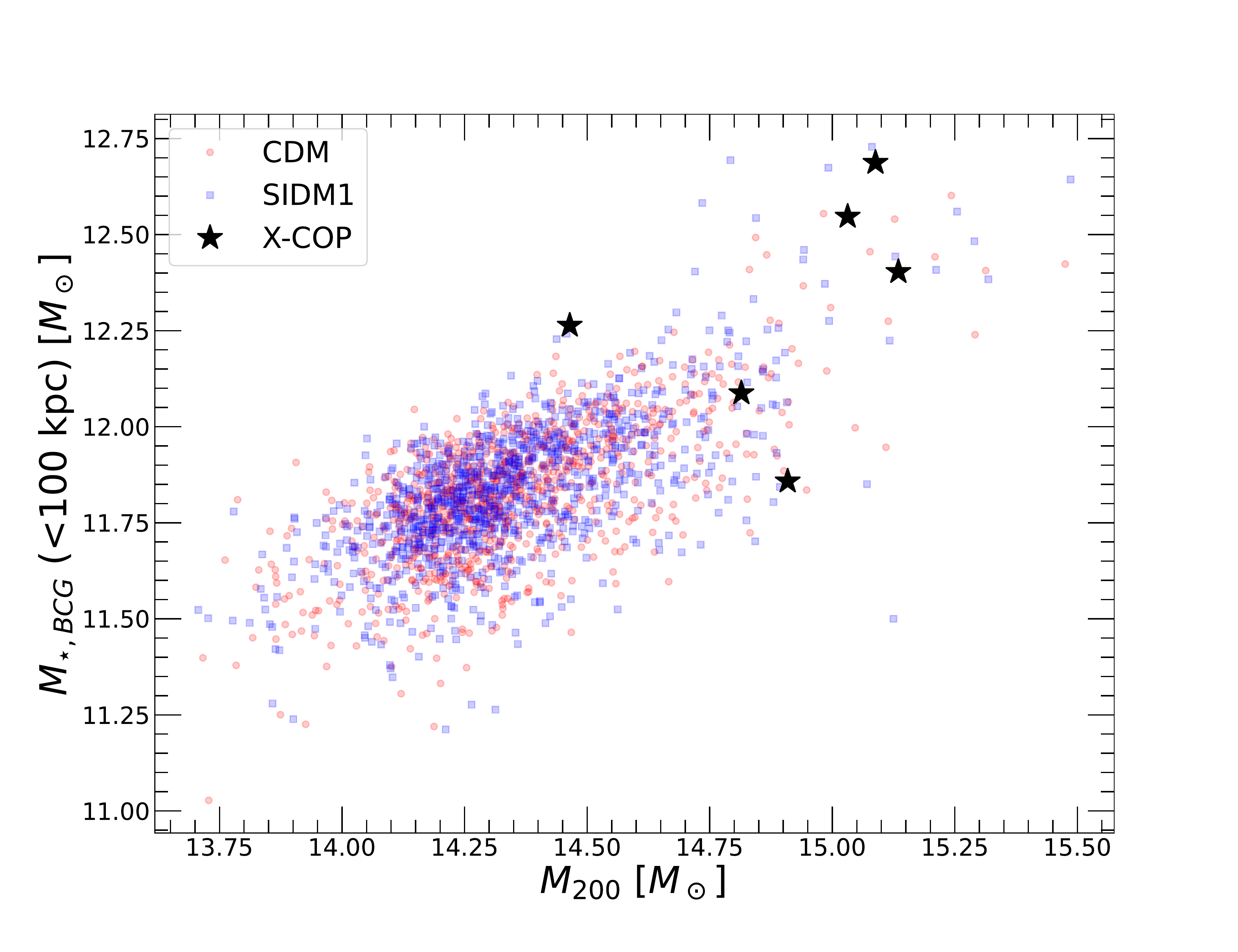}}
\caption{Impact of systematic uncertainties. \emph{Left:} Ratio between the value of $\alpha$ derived under the assumption of hydrostatic equilibrium ($\alpha_{HSE}$) and the true value of $\alpha$ when assuming two possible non-thermal pressure profiles (\citet{angelinelli20}, blue; \citet{nelson14}, orange) as a function of the input value of $\alpha$. \emph{Right:} Stellar mass of the BCG within 100 kpc in the \texttt{BAHAMAS} CDM (red) and SIDM1 simulations (blue) as a function of the halo mass. The black stars show the measured BCG stellar masses within the same aperture in X-COP clusters (see Paper \rom{1}).}
\label{fig:syst}
\end{figure*}

The constraints on the Einasto shape parameter presented in this paper were derived under the assumption that the gas is in hydrostatic equilibrium within the gravitational potential set by the DM (see Sect. \ref{sec:hydro}). While the comparison with weak lensing masses \citep{ettori19} and the measured gas fractions \citep{eckert19} indicate that the assumption of equilibrium is well justified in X-COP clusters, and the selection of the regular subsample based on the centroid shift excludes the most problematic cases, we do expect a certain level of non-thermal pressure to be present within the systems of interest. The relative importance of non-thermal pressure is expected to be non-uniform in the cluster's volume \citep[e.g.][]{rasia06,lau09,nelson14,biffi16,angelinelli20}, which affects the shape of the recovered mass profiles, and hence the Einasto shape parameter. 

To quantify the potential impact of hydrostatic bias on our results, we considered a fiducial cluster with a mass profile described by the Einasto functional form (Eq. \ref{eq:einasto}) and a typical gas density profile \citep{ghirardini19}. We computed its total pressure profile $P_{\rm tot}$ by numerically integrating Eq. \ref{eq:hse}, and we assumed that the measured thermal pressure $P_{\rm th}$ is a fraction of the total pressure,
\begin{equation}
P_{\rm th}(r) = P_{\rm tot}(r) - P_{\rm NT}(r)
\end{equation}
\noindent with $P_{\rm NT}(r)$ the non-thermal pressure, which is expected to be dominated by kinetic motions. We considered two possible cases: first, the non-thermal pressure profile predicted by \citet{nelson14}, which takes all the gas motions into account; second, the functional form introduced by \citet{angelinelli20}, which only selects the random, turbulent motions as a source of outward pressure. We then fitted the modified $P_{\rm th}$ profile with an Einasto model in the radial range $[0.01-1]R_{200}$ and compared the retrieved value of $\alpha$ with the input one. In the left-hand panel of Fig. \ref{fig:syst} we show the ratio of the measured valued of $\alpha$ in the HSE assumption ($\alpha_{\rm HSE}$) to the true input value ($\alpha_{\rm True}$) for a range of input values of $\alpha$. We can see that the HSE assumption typically overestimates $\alpha$ by $2-8\%$ depending on the assumed non-thermal pressure profile and the true value of $\alpha$, corresponding to differences of 0.01 for $\alpha_{\rm True}=0.18$. Therefore, we conclude that the hydrostatic assumption has little impact on the results presented here.

\section{Impact of the brightest cluster galaxy}
\label{sec:bcg}

Since measurements of the stellar mass profiles are currently available only for a subset of X-COP systems, in this paper we fitted an Einasto profile to the total gravitational field and compared the results to simulations including a detailed modeling of the baryonic components \citep[gas + stars, see][]{mccarthy17}. In case the modeling of baryonic processes implemented in \texttt{BAHAMAS} simulations does not accurately reproduce the observed baryonic mass profiles, the corresponding comparison between simulations and observations may be biased. In particular, the stellar content of the brightest cluster galaxy (BCG) dominates the gravitational field in the innermost regions, thus differences in the BCG mass profiles between simulations and observations could bias the comparison of the Einasto shape parameter. To verify this point, we determined the stellar masses of the BCGs in the simulation runs and compared them with the stellar masses of the BCGs in X-COP clusters \citep[see Paper \rom{1}]{loubser20}. In the right-hand panel of Fig. \ref{fig:syst} we show the BCG stellar masses within 100 kpc radius for the CDM and SIDM1 cases as well as the X-COP BCG stellar masses within the same aperture. We can see that the stellar masses of BCGs in the \texttt{BAHAMAS} model agree well with the observations and do not differ significantly from the CDM to the SIDM case, which shows that the comparisons presented in this paper are mildly impacted by the stellar content of the BCG.

While the analysis presented in the left-hand panel of Fig. \ref{fig:syst} does not show any potential issue associated with the BCG mass profile, it is worthwhile to quantify the potential impact of wrong BCG mass profiles on the Einasto shape parameter. Similar to the tests performed in Appendix \ref{sec:hse_alpha}, we considered a fiducial Einasto profile, this time fitted to the sum of BCG and DM profiles, and modified the stellar mass of the BCG by a conservative factor of 2 in the upward and downward directions. We then refitted the corresponding total mass profiles in the $[0.01-1]R_{200}$ range with a single Einasto profile and quantified the changes in $\alpha$ with respect to the input value. We found that $\alpha$ is anti-correlated with the BCG stellar mass, with larger values of the stellar mass corresponding to lower values of $\alpha$. For an input value $\alpha=0.18$ the retrieved value of $\alpha$ changes by $\pm0.02$, which is smaller than the statistical uncertainties in our measurements.

\end{document}